\documentclass[12pt]{article}
\usepackage[dvips]{graphicx}

\title{Onset of incommensurability in quantum spin chains}
\author{Kiyohide Nomura \\
Department of Physics, Kyushu University,
812-8581 Fukuoka, JAPAN
}

\begin{document}
\maketitle

\begin{abstract}
In quantum spin chains, it has been observed that the incommensurability occurs
 near valence-bond-solid (VBS)-type solvable points, and the correlation
 length becomes shortest at VBS-type points.  
Besides, the correlation function decays purely exponentially at VBS-type
 points, 
in contrast with the two-dimensional (2D) Ornstein-Zernicke type
behavior in other regions with an excitation gap. 
We propose a mechanism to explain the onset of the incommensurability and
the shortest correlation length at VBS-like points.
This theory can be applicable for more general cases.  
\end{abstract}


Frustration is one of the important problem in many-body physics.  
Recently, many things in frustrated systems become clearer 
with field theoretical approaches and numerical calculations (density
matrix renormalization group (DMRG) etc.).  
However, relatively little is known about commensurate-incommensurate
(C-IC) transitions induced by frustration in quantum spin systems.  
Among them, it is observed the C-IC change 
in the region with an excitation gap (therefore, it is not a {\em phase transition}) in
one-dimensional (1D) quantum spin models. 

For example, the C-IC change begins from the
Majumdar-Ghosh point \cite{Majumdar-Ghosh} in the S=1/2
next-nearest-neighbor (NNN) chain 
\begin{equation}
  H = \sum_j (S_{j} S_{j+1} + \alpha S_{j} S_{j+2}),
\label{eq:NNN-model}
\end{equation}
($\alpha_D=1/2$ is the solvable point), 
or from the valence-bond-solid (VBS) point \cite{Affleck-Kennedy-Lieb-Tasaki} 
in the S=1 bilinear-biquadratic chain,
\begin{equation}
  H = \sum_j (  S_{j} S_{j+1} + \alpha  (S_{j} S_{j+1})^2),
\label{eq:bilinear-biquadratic-model}
\end{equation}
($\alpha_D=1/3$ is the solvable point).

With a numerical diagonalization, Tonegawa and Harada observed in the S=1/2
NNN chain \cite{Tonegawa-Harada} 
that the incommensurability of the spin correlation begins near the
solvable point. 
However, from the DMRG calculation, Bursill et al. stated that in the static
structure factor (Fourier transform of spin-correlation function),
the C-IC change begins slightly different from the solvable point
for the S=1 bilinear-biquadratic chain \cite{Bursill-Xiang-Gehring},
and for the S=1/2 NNN chain \cite{Bursill-Gehring-Farnell-Parkinson-Xiang-Zeng}. 
In contrast, using the DMRG, Schollw\"ock, Jolicoeur, and Garel 
\cite{Schollwock-Jolicoeur-Garel} 
 have closely investigated the correlation
function of the S=1 bilinear-biquadratic chain
(\ref{eq:bilinear-biquadratic-model}), not the structure factor, 
and they found that the incommensurability begins right at 
the VBS point.  Adding this, they have pointed out the singular
behavior of the modulation wavenumber.

Secondly, it has been noticed that the correlation length is shortest at
the VBS-type point.  

Thirdly, at the VBS-type point, the spin-correlation function decays purely
exponentially, in contrast with 
the 2D (1+1D) Ornstein-Zernicke type (that is, 2D Green function for the free massive
particle) behavior in the other region with a gap. 

Although Schollw\"ock et al. have tried to explain these features on the
analogy of the 1D classical frustrated Ising spin with finite
temperature, several features are
different from the quantum spin chain case, which we intend to explain
in this study. 
After that, Kolezhuk, Roth, and Schollw\"ock applied their idea to the
S=1 NNN chain \cite{Kolezhuk-Roth-Schollwock96, Kolezhuk-Roth-Schollwock97},
where they called the shortest correlation length point as the
``disordered point'', generalizing the VBS-type point with the above
features.  

Recently, F\'ath and S\"ut\~o proposed an effective field theory \cite
{Fath-Suto}
to explain these features in the C-IC change of the S=1 bilinear-biquadratic
chain. Although their results seem natural, they have not derived
their effective field theory directly from the lattice Hamiltonian.  
And from their theory, the C-IC change near the Majumdar-Gosh point
cannot be explained.  
Therefore, it is meaningful to construct a theory on firmer ground. 
Considering the analyticity of the structure factor $S(q)$ around the
disordered point, where several exact results are known, we propose another
approach to this problem.  Since, what we observe experimentally is not
Hamiltonian itself, but structure factors or other quantities, thus
our approach may be more natural in a physical sense.

\begin{figure}[here]
 \includegraphics[width=60mm]{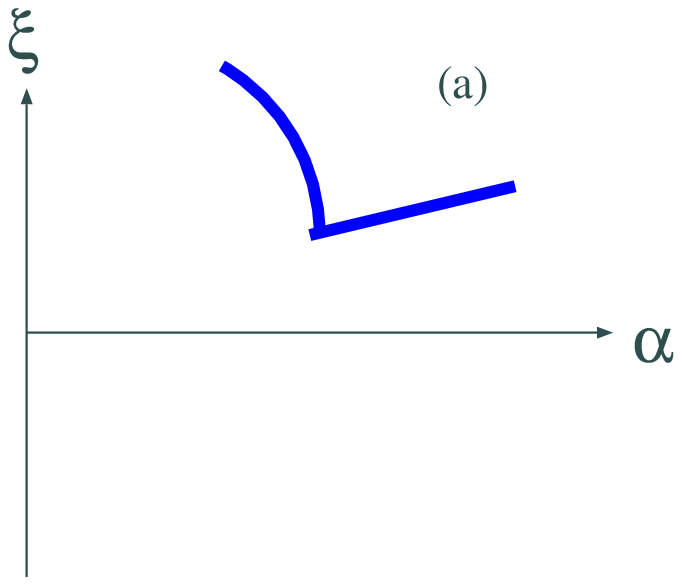}
 \hskip 10mm
 \includegraphics[width=60mm]{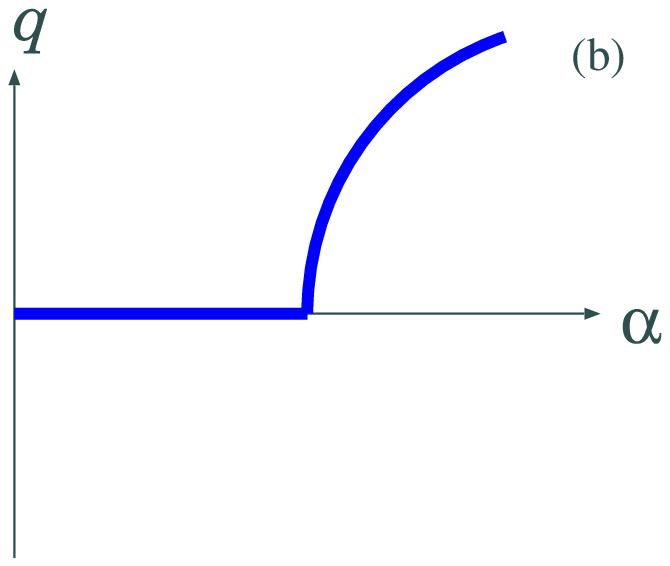}
 \caption{Singular behaviors of correlation length (a) and wave number (b).}
\label{fig:correlation-length-wave-number}
\end{figure}

We summarize the results by Schollw\"ock et al. as (see also
Fig.\ref{fig:correlation-length-wave-number})
\begin{enumerate}
 \item In the commensurate region ($\alpha < \alpha_D$), the correlation
 function behaves as the 2D Ornstein-Zernicke-type 
       \begin{equation}
	\langle S_0 S_n \rangle \approx (-1)^n |n|^{-1/2} \exp (-|n| / \xi(\alpha)),
       \end{equation}
       where far enough from the disordered point $\alpha_D$.
 \item In the incommensurate region ($\alpha > \alpha_D$) ,
       \begin{equation}
	\langle S_0 S_n \rangle \approx (-1)^n \cos (q (\alpha) n)
	|n|^{-1/2} \exp (-|n| / \xi(\alpha)) ,
       \end{equation}
       where far enough from the disordered point $\alpha_D$.
 \item At the disordered point $\alpha_D$,
      \begin{equation}
	\langle S_0 S_n \rangle \approx (-1)^n \exp (-|n| /
	\xi(\alpha_D)) .
       \end{equation}
 \item From the commensurate side,
       \begin{equation}
	\frac{d \xi (\alpha)}{d \alpha}|_{\alpha_D} = - \infty .
       \end{equation}
 \item From the incommensurate side,
       \begin{equation}
	\frac{d q (\alpha)}{d \alpha}|_{\alpha_D} = \infty .
       \end{equation}
 \item From the incommensurate side, by fitting as
       \begin{equation} 
	q (\alpha) \propto (\alpha -\alpha_D)^{\sigma} ,
       \end{equation}
       we obtain $\sigma \approx 1/2$.
 \item
       Neither the $\exp(-|n|/\xi)$ nor the $|n|^{-1/2} \exp(-|n|/\xi)$
       form cannot describe the correlation function near the disordered
       point $\alpha_D$.  
\end{enumerate}
Note that, strictly speaking, the 2D Ornstein-Zernicke form is 
the modified Bessel function \cite{Nomura}, 
\begin{equation}
G (\vec{n} ) =
 \int \frac{\exp(i \vec{q} \cdot \vec{n}) d^2 q}{\vec{q}^2 +\xi^{-2}} =
 2 \pi K_0(|n|/\xi) ,
\end{equation} 
which behaves $|n|^{-1/2} \exp (- |n| / \xi) $ in large distance. 
And static correlation function is 
\begin{equation}
 G(0, n_1) = \pi \int \frac{\exp(i q_1 n_1)}{\sqrt{q_1^2
 +\xi^{-2}}} d q_1 .
\end{equation}

To explain these features, we consider the
static structure factor in the complex $q$ plain, 
\begin{equation}
 S(q) \equiv \sum_n \exp (i q n) (-1)^n \langle S_0 S_n \rangle .
\end{equation}
The simple minded Fourier transforms of the correlation functions are
\begin{enumerate}
 \item 
At the disordered point,
\begin{equation}
 S(q) \propto \frac{1}{q^2 + m^2}.
\end{equation}
\item
In the commensurate region,
\begin{equation}
 S(q) \propto \frac{1}{\sqrt{q^2 + m^2}}.
\end{equation}
\item
In the incommensurate region,
\begin{equation}
 S(q) \propto \frac{1}{\sqrt{(q- q(\alpha))^2 + m^2}}
 + \frac{1}{\sqrt{(q + q(\alpha))^2 + m^2}}.
\end{equation}
\end{enumerate}
However, with the above forms, one cannot connect these three region.
In fact, in terms of the singularity in the complex $q$ plain, 
there are poles at the
disordered point, in contrast with the branch-cuts in the other
(commensurate, incommensurate) regions.

In order to unify these three region, we reconsider the relation 
between the pole and the branch-cut.
A pole and a branch-cut can be deformed each other as following,
\begin{equation}
 f(z) \equiv (z^2 -d)^{-1/2},
\end{equation}
which has two branch-points at $z= \pm \sqrt{d}$ for  $d>0 $,
a pole at $z=0$ for  $d=0 $, and 
two branch-points at $z= \pm i \sqrt{|d|}$ for  $d < 0 $.  
Note that $f(z)$ is an odd function $f(-z) = - f(z)$, considering
the analytic continuation around the branch-cut.  

\begin{figure}[here]
 \includegraphics[width=45mm]{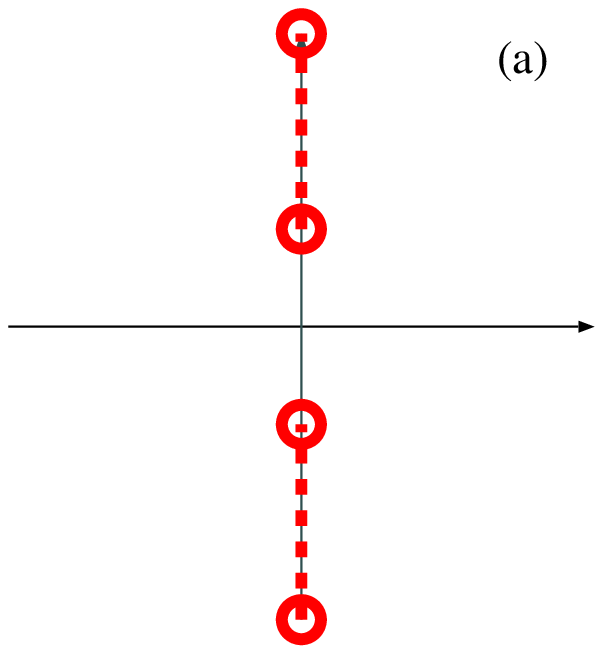}
 \hskip 5mm
 \includegraphics[width=45mm]{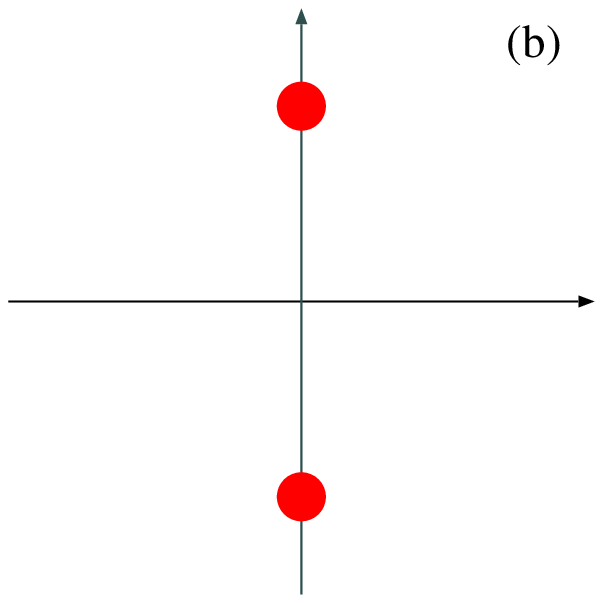}
 \hskip 5mm
 \includegraphics[width=45mm]{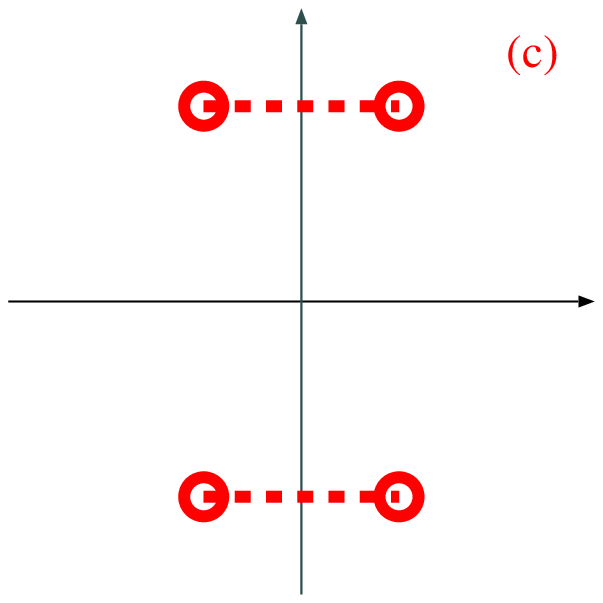}
 \caption{Continuous deformation of branch-cuts and poles.
 They correspond to commensurate region (a),
 disordered point (b), incommensurate region (c),
}
 \label{fig:singularities}
\end{figure}

Besides, to relate it with physical phenomena, we consider 
the following assumptions:
\begin{enumerate}
 \item $S(q)$ is analytic except several {\em algebraic} singular points. 
 \item $S(q)$ is real on the real axis $(S (q) = \bar{S}(\bar{q}))$. 
 \item From parity, $S(q)=  S (- q)$.
\item $S(q)$ is varied continuously with a parameter $\alpha$.
\item At the disordered point $\alpha_D$, $S(q)$ is described with two poles.
\item Singular points and branch-cuts should not cross the real axis.
\end{enumerate}
From these assumptions, we can obtain the simplest singularity structure as 
Fig. \ref{fig:singularities}, 
that is, there are two poles symmetric
with real axis, or four branch-points continuously deformable from
them. 
The static structure factor, originated from the singular terms, is
given as 
\begin{eqnarray}
S_{sing}(q) &=& A f(q + \tilde{m} i ) f(q - \tilde{m} i ) \nonumber \\
 &=& A [q^4 + 2(\tilde{m}^2-d) q^2 + \tilde{m}^4 + d^2 + 2\tilde{m}^2 d]^{-1/2},
\label{eq:structure-factor}
\end{eqnarray}
where real parameters $A, \tilde{m}, d$ may depend on $\alpha$, especially 
$d \propto \alpha- \alpha_D$. 
Although this result for the static structure is the same as 
in \cite{Fath-Suto}, our theory is based on the analyticity of $S(q)$ 
and the exact $S(q)$ at the VBS point.
Note that without the assumption for
{\em algebraic } singularities, there remain other possibilities 
\begin{eqnarray}
S_{sing}(q) &=&
A \frac{i}{2 \tilde{m}}[ f(q + \tilde{m} i ) - f(q - \tilde{m} i )] \nonumber\\
&=&  \frac{A}{\tilde{m}}\frac{1}{\sqrt{(q^2 - \tilde{m}^2 -d)^2 +4 q^2 \tilde{m}^2}}
 \cos \left( \frac{1}{2} \arctan (\frac{2 q \tilde{m}}{q^2 -\tilde{m}^2 -d})\right),
\label{eq:structure-factor2} \\
S_{sing}(q) &=&
A \frac{1}{2 q}[ f(q + \tilde{m} i ) + f(q - \tilde{m} i )].
\label{eq:structure-factor3}
\end{eqnarray}
The assumption for {\em algebraic} singularities physically relates 
to {\em the locality of interactions}\cite{NB1}.

Although it seems reasonable four branch-points in the incommensurate region, 
how do we interpret four branch-points in the commensurate region?
In the large distance, or the small wave number $q$, to $S(q)$ on the real axis 
mainly contribute the two branch-points closer to the real axis, 
thus it seems only two branch-points when 
$
0 < 1 - |d|/ \tilde{m}^2 \ll 1
$
.

When we set $d \approx c_1 (\alpha - \alpha_D); \; c_1>0 $ 
in eq. (\ref{eq:structure-factor}), 
we can explain systematically the singular behaviors of 
$\xi(\alpha),q(\alpha)$ near the disordered point $\alpha_D$. And we can calculate the
analytic form for $S(q)$ in the neighborhood of $\alpha_D$.  
In addition, near $\alpha_D$, $\tilde{m}$ may behave as
$ \tilde{m} \approx \tilde{m}_0 + c_2 (\alpha-\alpha_D)$.
It should be $c_2<0$, since the correlation length increases in the
incommensurate region. 

Generally, the requirement for the amplitude is
$A \neq 0$, since for $A = 0$ the
correlation function becomes perfectly zero.  
However, at the Majumdar-Ghosh point in the S=1/2 NNN chain, the correlation
function behaves as 
\begin{equation}
 \langle S_0 S_n \rangle = 0,  \; (for |n| \ge 2 ),
\end{equation}
which can be interpreted as the $A (\alpha) \propto (\alpha -\alpha_D)$
case or the $A (\alpha) \propto (\alpha -\alpha_D)^2$ case.  
The latter possibility comes from that spin variables may depend as
$S_n \propto (\alpha -\alpha_D)$.

Until now, we have neglected the lattice
structure and the terms which are regular in the whole complex $q$ plain.  
To include the lattice structure, we should require 
$S(q) = S(q+ 2\pi) $.  
Thus, the full static structure factor is 
\begin{equation}
 S(q) = \sum_{j= -\infty}^{\infty} S_{sing} (q+ 2 \pi j) + S_{reg}(q),
\end{equation}
where
\[
 S_{reg}(q) \equiv \sum_{j= -\infty}^{\infty} a_j \cos (j q). 
\]
Note that to calculate the spin-correlation function, 
it is enough to consider
the singularities of $S(q)$ within the Brillioun zone, 
because of periodicity.  

In summary, the C-IC change can be regarded as a fusion-fission of
singularities in the complex $q$ plain.  The Lifshitz point, 
where the static structure factor changes from a single peak to double
peaks, is a natural consequence of this fusion-fission of singularities 
\cite{Schollwock-Jolicoeur-Garel,Kolezhuk-Roth-Schollwock96,Kolezhuk-Roth-Schollwock97}
(see in (\ref{eq:structure-factor}), 
\begin{equation}
 \frac{\partial^2 S_{sing}(q)}{\partial q^2}(q=0) \propto A (\tilde{m}^2 -d)
\end{equation}

).

The argument in this paper should be checked in the DMRG calculation near
the disordered point.  
After checking the expectation for the static structure factor, one
can extend it to the dynamical structure factor.  
Since the VBS-type states \cite{Affleck-Kennedy-Lieb-Tasaki} are obtained
not only in the 1D quantum system, 
but also in the 2D and 3D  quantum systems, it is interesting whether such
C-IC changes may occur at the VBS-type points in higher dimensions.

\end{document}